%% file: book.tex
\RequirePackage{etoolbox}
\csdef{input@path}%
{%
 {sty/},
 {img/},
}
\documentclass{elsevierbook}

\usepackage{natbib}
\usepackage{url}

\usepackage{lineno}

\begin{document}

\Frontmatter

\Mainmatter

\include{chapter7-TNOs}


\Backmatter
\end{document}

%% file: chapter7-TNOs.tex
\begin{frontmatter}
\setcounter{chapter}{6}
\chapter{Machine Learning Assisted Dynamical Classification of
Trans-Neptunian Objects}\label{chap1}

Authors: Kathryn Volk (Planetary Science Institute) and Renu Malhotra (The University of Arizona)

\begin{abstract}
Trans-Neptunian objects (TNOs) are small, icy bodies in the outer solar system. 
They are observed to have a complex orbital distribution that was shaped by the early dynamical history and migration of the giant planets. 
Comparisons between the different dynamical classes of modeled and observed TNOs can help constrain the history of the outer solar system. 
Because of the complex dynamics of TNOs, particularly those in and near mean motion resonances with Neptune, classification has traditionally been done by human inspection of plots of the time evolution of orbital parameters. 
This is very inefficient. 
The Vera Rubin Observatory’s Legacy Survey of Space and Time (LSST) is expected to increase the number of known TNOs by a factor of ~10, necessitating a much more automated process.
In this chapter we present an improved supervised machine learning classifier for TNOs.
Using a large and diverse training set as well as carefully chosen, dynamically motivated data features calculated from numerical integrations of TNO orbits, our classifier returns results that match those of a human classifier 98\% of the time, and dynamically relevant classifications 99.7\% of the time.
This classifier is dramatically more efficient than human classification, and it will improve classification of both observed and modeled TNO data.

\end{abstract}

\begin{keywords}
\kwd{minor planets}
\kwd{TNOs}
\kwd{orbital resonances}
\end{keywords}

\end{frontmatter}

\section{Introduction to Dynamical Classification of TNOs}\label{s:intro}

Transneptunian objects (TNOs) are small solar system bodies with semimajor axes ($a$) in the range 30--2000~au, beyond Neptune but interior to the Oort Cloud. 
Their orbits are perturbed by Neptune and other solar system giant planets but relatively unperturbed by external forces such as those of passing stars and galactic tides. 
Astronomical surveys for TNOs have thus far sampled only a very small fraction of the whole population.
The observed set of approximately 4000 TNOs have complex distributions in $a$, eccentricity ($e$), and inclination ($i$) that reveal multiple dynamical sub-classes (described in Section~\ref{s:classes}).
The distribution of TNOs in these sub-classes has revealed important details about the dynamical history of the outer solar system's giant planets and early planetesimal disk, though many open questions remain (see, e.g., a recent review by \citealt{Gladman:2021}).
The Vera Rubin Observatory's Legacy Survey of Space and Time (LSST) is expected to increase the number of known TNOs to $\sim40,000$ \citep{Ivezic:2019,Schwamb:2019}. 
We have thus far heavily relied on manual dynamical classification of TNOs, but leveraging the full LSST TNO dataset to further constrain dynamical models of the early solar system will require automated approaches.

The rest of Section~\ref{s:intro} provides an overview of the dynamical classes of TNOs, the challenges faced in classifying TNOs, and the need for improved machine learning classifiers.
Section~\ref{s:classifier} describes an improved classifier based on a supervised learning approach with a large and diverse labeled training/testing dataset.
Section~\ref{s:conclusions} summarizes our work and describes anticipated future applications.

\subsection{The dynamical classes of TNOs and current approaches to classification}\label{s:classes}

\begin{figure}
    \centering
    \includegraphics[width=4in]{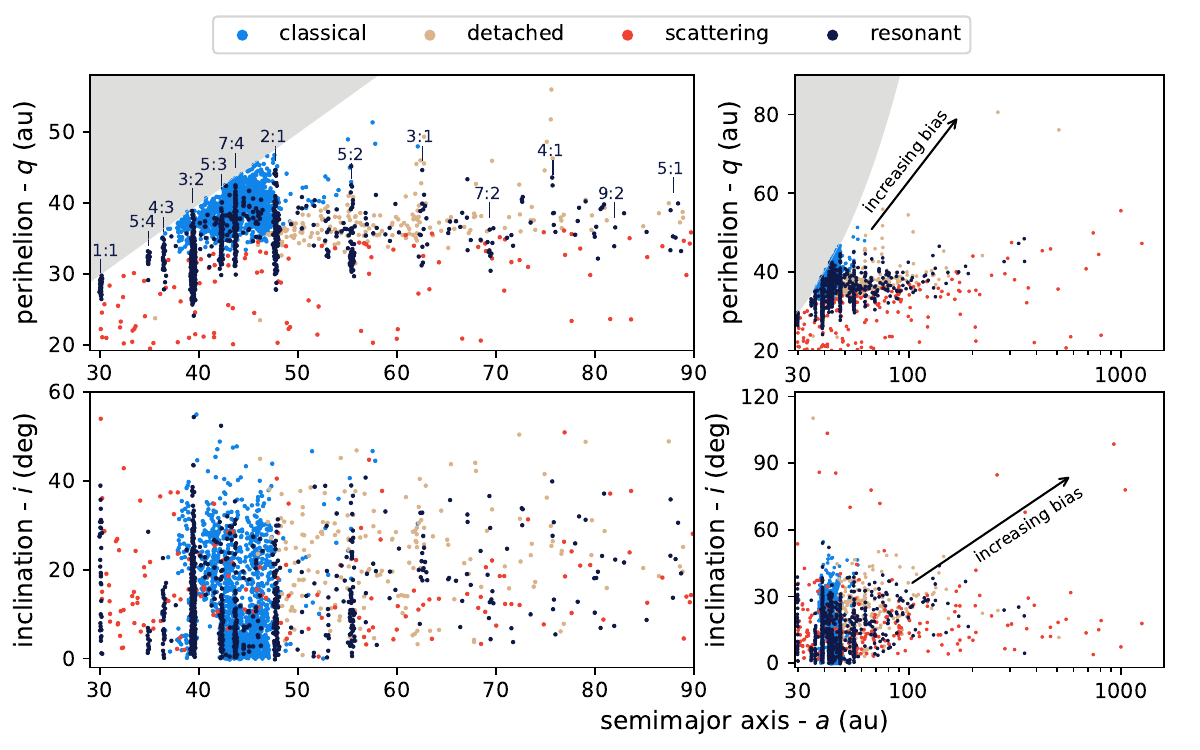}
    \caption{Current inventory of 3357 multi-opposition TNOs with orbits sufficiently well-constrained to classify using the \cite{Gladman:2008} scheme. 
    The top panels show perihelion distance vs semimajor axis (the grayed out areas are unphysical) while the bottom panels show ecliptic inclination vs semimajor axis. 
    The right panels show a zoomed-out view over a larger parameter space range; note the log-scale for semimajor axis.
    The left panels show a zoomed-in view of the closer-in TNO populations; various prominent mean motion resonances with Neptune are labeled.
    The data underlying this plot is published in \cite{Volk:2024}, and a version of this plot with a smaller sample of TNOs was published in \cite{Gladman:2021}.  }
    \label{f:inventory}
\end{figure}

Figure~\ref{f:inventory} shows the current census of TNOs divided into four dynamical classes by \cite{Volk:2024}: resonant, scattering, classical, and detached.
We describe the exact definitions of these classes below, but it is useful to review their broad features and implications.
The resonant TNOs are objects that librate in Neptune's exterior mean motion resonances (described in detail in Section~\ref{ss:resonances}).
They are prominent both in the biased observational sample \citep[e.g.][]{Elliot:2005,Kavelaars:2009,Bannister:2018,Bernardinelli:2022,Smotherman:2023} as well as intrinsically \citep[e.g.][]{Gladman:2012,Adams:2014,Volk:2016,Chen:2019,Crompvoets:2022}.
The present-day large stable resonant TNO populations imply the capture of TNOs into Neptune's resonances during the epoch of giant planet migration (see, e.g., \citealt{Malhotra:1995,Malhotra:2019}). 
The resonant TNOs are thus of particular interest to identify because they provide an important test for dynamical models (discussed in \citealt{Gladman:2021}).
The classical TNOs appear in Figure~\ref{f:inventory} as a collection of relatively lower-$e$ (higher perihelion distance $q$) orbits concentrated at semimajor axes between Neptune's 3:2 and 2:1 resonances. 
The lowest-$e$ and lowest-$i$ members of the classical belt (often called `cold' classicals) are thought to be the only in-situ remnant of the outer solar system's original planetesimal disk, while the more dynamically excited `hot' classical population was transplanted into this region from closer-in portions of the disk that was dispersed during planet migration (see discussion in recent reviews by \citealt{Morbidelli:2020} and \citealt{Gladman:2021}).
The scattering TNOs are objects with low-enough perihelion distances that they experience strong perturbations from Neptune that can change their orbital energy (and thus semimajor axis) on timescales much shorter than the age of the solar system.
In Figure~\ref{f:inventory}, only resonant TNOs have perihelion distances that overlap with the scattering population, but they are phase protected from encountering Neptune at perihelion by their resonant orbits.
Detached TNOs have larger $q$ than the scattering population and correspondingly more stable orbits; these are TNOs outside the classical belt region that are neither resonant with Neptune nor experiencing significant changes in $a$ due to planetary encounters.
Both the scattering and detached TNOs are remnants of the original planetesimal disk that was scattered outward when the giant planets migrated to their present orbits (see, e.g., \citealt{Gomes:2008,Dones:2015}).
The distributions of TNOs amongst these different dynamical classes provide critical constraints on the exact details of the planet migration era. 
These distributions are affected by many observational biases that must be accounted for when comparing observations to theoretical models. 
The first step in this comparison is dynamical classification of the observed orbits.

In this chapter, we will define TNO dynamical classes using the scheme presented by \cite{Gladman:2008}.
In this nomenclature system, the observed orbit of a TNO is numerically integrated under the influence of the Sun and all four giant planets for 10 Myr and the integration is used to classify the TNOs as follows: 
\begin{itemize}
    \item Resonant TNOs: the resonant angle for one of Neptune's external $p$:$q$ mean motion resonances shows libration for at least 50\% of the integration
    \item Scattering TNOs: a non-resonant orbit that exhibits a barycentric semimajor axis change $\Delta a> 1.5$~au during the integration
    \item Detached TNOs: a non-resonant, non-scattering orbit with $e>0.24$ (which tends to coincide with $a\gtrsim50$~au)
    \item Classical TNOs: a non-resonant, non-scattering orbit with $e\leq0.24$ (these are most concentrated between Neptune's 3:2 and 2:1 resonances; $39.4 < a < 47.8$~au).
\end{itemize}
\cite{Gladman:2008} further divide the classical TNOs into inner, main, and outer based on semimajor axis, but we do not consider these divisions here. 
Note that the order of operations in implementing this scheme is important.

The dynamical behavior of TNOs is a spectrum rather than being discrete, so the above scheme represents just one set of possible dividing lines, with a number of specific choices to divide the spectrum of dynamical evolution.
The first choice is the timescale over which the dynamical behavior is classified. 
For the majority of observed TNOs, 10 Myr is a timescale that will capture a reasonable picture of their present dynamics.
Resonant libration periods are typically $10^4-10^5$~years, so this timescale covers many libration cycles.
For close-in TNOs ($a\lesssim50$~au), secular variations in $e$ and $i$ have timescales of a a few Myr, so 10 Myr will capture their full range.
For large-$a$ TNOs (especially those with $a\gtrsim100$~au), 10 Myr is no longer sufficient to capture this range. 
As we discover more large-$a$ TNOs, it will be worth reconsidering appropriate classification timescales, because longer-term $e$ variations can  cause objects to switch between the scattering and detached populations.

The next specific choice is the
$\Delta a>1.5$~au definition for  the scattering population. 
This is a reasonable dividing line between stable and unstable orbits for the moderate-$a$ ($a\lesssim100$~au) scattering population that dominates the observed set of TNOs (scattering objects with smaller $a$ are more likely to be observed than those with larger $a$ in brightness limited surveys).
Some unstable classical objects do not meet this threshold to qualify as scattering objects because their tightly-bound orbits require larger energy changes, and some very distant objects with large perihelia would likely be better described as diffusing due to smaller perturbations on their weakly bound orbits \citep[e.g.][]{Bannister:2017}.
This is also a case for which changes in the criterion, such as defining a relative $\Delta a/a$ threshold and/or adding an additional class of diffusing objects, should be considered in the future as more TNOs are discovered.

The division between the classical and detached TNOs might similarly evolve with more discoveries.
The motivation for the eccentricity cut between these populations is to accommodate the possibility of a second or extended low-eccentricity classical belt beyond the currently known one. 
A more distant population of TNOs on nearly circular orbits has not yet been ruled out by observations (see discussion in \citealt{Gladman:2021}), and currently unseen TNO populations have been proposed based on dust measurements \citep[e.g.][]{Doner:2024}.  
If discovered, such a population would likely not have the same dynamical emplacement history as the known detached population, a motivation for providing a separation.
However, for the known TNOs, the classical and detached populations share very similar dynamics even if they have different semimajor axis ranges. 
We will test our classifier with their current, separate class labels as well as combining them into a single class in Section~\ref{ss:performance}.

The last choice of note in the scheme above is the 50\% threshold for resonant libration over 10 Myr.
This places many resonance-interacting objects in the classical and detached populations and many intermittently resonant objects in the resonant population. 
We will discuss the resonant TNOs in detail in the next section, including the challenges presented by TNOs with intermittent resonant libration.

Thus far, the most accurate way to divide TNOs into the four classes described above is to rely heavily on manual inspection of the 10 Myr integration outputs. 
This is essentially entirely due to how difficult it can be to unambiguously identify resonant behavior in automated time-series analyses.
While scattering objects are relatively easy to identify through their large changes in $a$, it is more difficult to classify the objects that remain relatively stationary in $a$ over long timespans. 
A lack of mobility in $a$ can mean an object is not experiencing significant perturbations (i.e., a classical or detached object), or it can mean an object is resonant with Neptune; 
typically, resonant angles are calculated and checked for libration to identify resonant objects.
However, as we describe in the next section, determining resonant status based solely on the analysis of resonant angles is both challenging and inefficient. 
In contrast, it is remarkably easy to visually identify most resonant behavior by examining plots of $a$ vs time.
Humans are very good at pattern recognition, and after examining thousands of plots of TNO orbits, a human finds it very easy to distinguish resonant librations in $a$ from non-resonant variations in $a$.
This visual inspection combined with simple codes that can apply the other criteria above result in the most accurate classifications; but this approach is not scalable to very large numbers of TNOs.

\subsection{The inherent challenges of identifying resonant TNOs}\label{ss:resonances}

Neptune's external mean motion resonances have resonant angles with the following form:
\begin{equation}
    \phi = p\lambda - q\lambda_N - m\varpi - r\varpi_N - n\Omega - s\Omega_N,
\end{equation}
where $\lambda$ is the mean longitude, $\varpi$ is the longitude of perihelion, and $\Omega$ is the longitude of ascending node; in each case the subscript N refers to the orbit of Neptune and the non-subscripted elements are for the TNO.
The integers $p$ and $q$ ($p>q>0$) describe the period ratio of the TNO and Neptune such that the TNO completes $q$ orbits for every $p$ orbits of Neptune.
The integers $m$, $r$, $n$, and $s$ (all $\ge0$) must sum to equal $p-q$, and the integers $n$ and $s$ must be zero or even.
The value of $p-q$ is often referred to as the `order' of a resonance, because in classical analyses of mean motion resonances of low-$e$, low-$i$ orbits, the strength of the perturbation associated with a particular resonant argument is proportional to $e$ and/or $\sin i$ raised to the power of their corresponding integers $m$, $r$, $n$, and $s$ (see, e.g., \citealt{ssdbook} for a full discussion of resonant angles in the context of the disturbing function).
The eccentricities and inclinations of TNOs are typically much larger than those of Neptune, so resonances with $r,s\ne0$ are much weaker than those with $m,n\ne0$.
We will thus only be considering a simplified range of Neptune's resonances in this chapter:
\begin{equation}\label{eq:phi}
    \phi = p\lambda - q\lambda_N - m\varpi - n\Omega.
\end{equation}
We refer to resonances with $n=0$ as eccentricity-type resonances and term those with $n\ne0$ as mixed-type resonances.
In principle, inclination-type resonances ($m=0$) can also occur, but there are not yet any observed TNOs confirmed in such resonances; in our entire training/testing set of orbit integrations (Section~\ref{ss:training-set}), there is only one case for which the inclination-type resonant argument librates while the mixed and eccentricity-type arguments do not.
Eccentricity-type resonances are, by far, the most common resonances in the observed set of TNOs (see Table~\ref{t:training-set}).

An immediate challenge presented in the search for resonant TNOs is that there are an infinite number of possible combination of integers $p$ and $q$ that could describe an object's period ratio with Neptune.
We must somehow limit that infinite set.
As noted above, for low-$e$, low-$i$ orbits, one could limit the choices of $p$ and $q$ by some maximum difference between the integers based on a resonance strength argument.
However, the eccentricities of many TNOs are so large that the traditional $p-q$ resonance order is not particularly useful for predicting the strength of an eccentricity-type resonance.
The strength of a resonance is only predicted by $p-q$ if all of the conjunctions between the two resonant bodies are dynamically meaningful.
Bodies with a period ratio of $p$:$q$ will experience $p-q$ conjunctions over one resonant cycle (the time it takes to return to their starting configuration, $q$ orbits of the TNO).
For low-$e$ orbits, the separation between the two bodies at conjunction does not vary hugely over the resonant cycle, so each conjunction is a significant perturbation; having many conjunctions over a resonant cycle tends to `smear' out the effects of the resonance, making it dynamically irrelevant for large-enough $p-q$.
For the relatively lower-$e$ TNOs in the classical belt region ($a\lesssim50$~au), the low $p-q$ resonances do dominate in strength (see, e.g., \citealt{Gallardo:2006atlas}); though some observed TNOs do display libration in the classical belt region resonances with $p-q>10$.

However, when we consider a more distant resonant TNO on a highly-eccentric orbit with a large semimajor axis, $p-q$ becomes a less useful proxy for resonance strength. 
The vast majority of such a resonant TNO's conjunctions with Neptune will occur when the TNO is far from perihelion and at such large separations that they have very little effect on the TNO's orbit.
Only the interactions between the TNO and Neptune when the TNO is at or near perihelion can happen at small-enough separations that they can affect its evolution. 
Thus, for high-$e$ TNO orbits, large values of $q$ (the number of TNO perihelion passages over one resonant cycle) rather than large $p-q$ will cause Neptune's perturbations to `smear' out and make the resonance weaker. 
The resonances with smaller values of $q$ will be stronger regardless of $p-q$ (see, e.g., \citealt{Gallardo:2006atlas,Lan:2019}).
We will hereafter refer to $q$ as the resonance order when considering Neptune's resonances (following, e.g., \citealt{Pan:2004}).
It is the best proxy for resonance strength for high-$a$ TNOs, and because of the radial distribution of low-$e$ TNOs, the resonances relevant for low-$e$ orbits that have large $p-q$ also have large $q$; so they are high-order resonances in either definition of ``resonance order".

\begin{figure}
    \centering
    \includegraphics[width=4in]{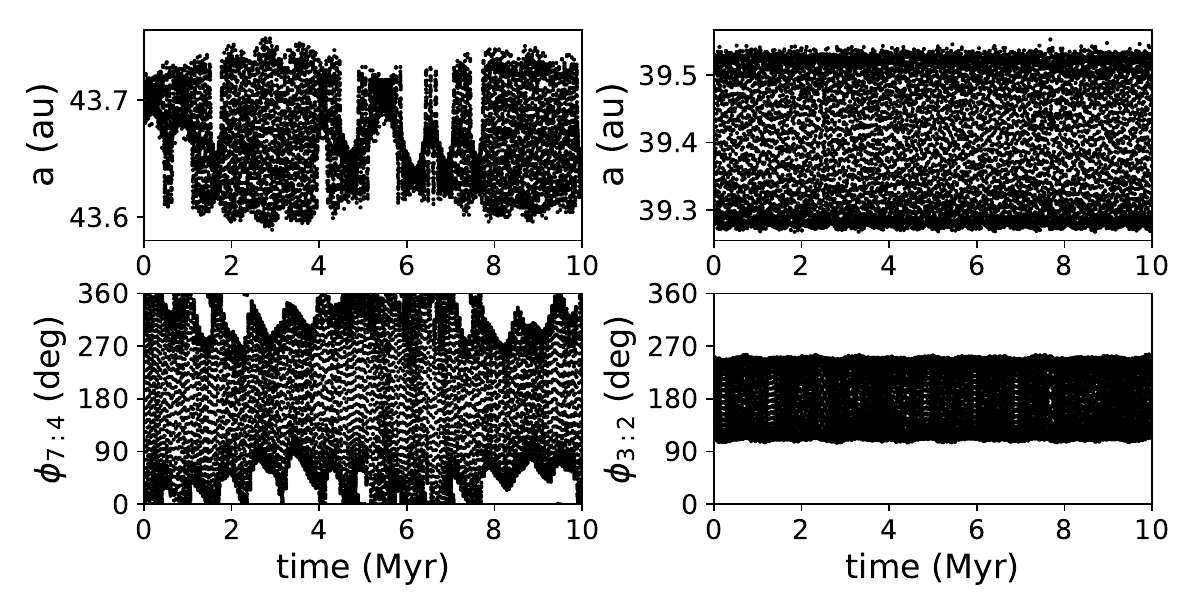}
    \caption{Evolution of semimajor axis (top panels) and resonant angles (bottom panels) of the best-fit orbits of TNO 2013 UG17 in Neptune's 7:4 resonances (left panels) and TNO 534161 in Neptune's 3:2 resonances (right panels).
    The resonance angles are $\phi_{7:4} = 7\lambda - 4\lambda_N - 3\varpi$  and $\phi_{3:2} = 3\lambda - 2\lambda_N - \varpi$.
    TNO 534161's evolution is an example of clean, relatively low-amplitude resonant libration that is easily characterized using simple bounds on $\phi$. 
    TNO 2013 UG17's evolution is a very typical example of intermittent libration that is more difficult to characterize, but its resonant nature is very readily recognized by an experienced human inspecting the evolution of $a$. }
    \label{f:messy-res}
\end{figure}

The above considerations still leave us with the question: 
what cutoff value of $p-q$ and/or $q$ should we adopt for the practical task of dynamical classification?
We will not have a conclusive answer to this question in this chapter, but the development and performance of our machine learning classifier (detailed in Section~\ref{s:classifier}) offers some insights. 
While all of the resonant TNO classifications (described in Section~\ref{ss:training-set}) in the training and testing dataset for the classifier are based on visual confirmation of a librating resonant angle (the check of which is often triggered by visual inspection of the semimajor axis evolution), we do not feed any information about resonant angles to the machine learning classifier.
This is because of the difficulties described above with the number of potential resonant angles that would need to be calculated and tested for libration. 
Even if we limit ourselves to a set of $p$:$q$ resonances with previously identified members, the number of resonant angles in the classical belt region can easily become impractically large to calculate and check (every variation of $m$ and $n$ in equation~\ref{eq:phi} must be considered for every $p$ and $q$ combination). 
An arbitrary limit would be premature in the face of the many new TNO discoveries to come that might occupy new resonances. 
To construct our labeled training and testing set, we did in fact calculate a \textit{very} large number of potential resonant angles to very high resonance orders for every particle and perform an automated simple running-window analysis to assess if any of those angles librated (i.e. were confined to a range of $\Delta \phi<355^\circ$).
This extensive computer-based search often took more computational time than the 10 Myr orbital integration, and it often missed borderline cases of intermittent, or merely very high-amplitude resonant libration. 
These missed resonant cases were almost always immediately apparent to the human eye based on visual inspection of plots of the semimajor axis time series for the particle. 
In these cases, a manual search for a librating resonant angle was performed to confirm resonance.
A typical example is shown in the left panels of Figure~\ref{f:messy-res}, where the resonant nature of the orbit is very apparent to an experienced human from the time evolution of $a$, but the libration of the resonant angle is more difficult to characterize using any simple computer-based analyses such as testing for confinement in a series of time windows; the right panels in Figure~\ref{f:messy-res} show an example of well-behaved libration for comparison.
While the $\phi$ evolution could be characterized with more complex analyses (including machine learning approaches!), doing such analyses for a large enough set of potential resonant angles for any given particle becomes computationally prohibitive.

\begin{figure}
    \centering
    \begin{tabular}{c c}
    \includegraphics[width=2in]{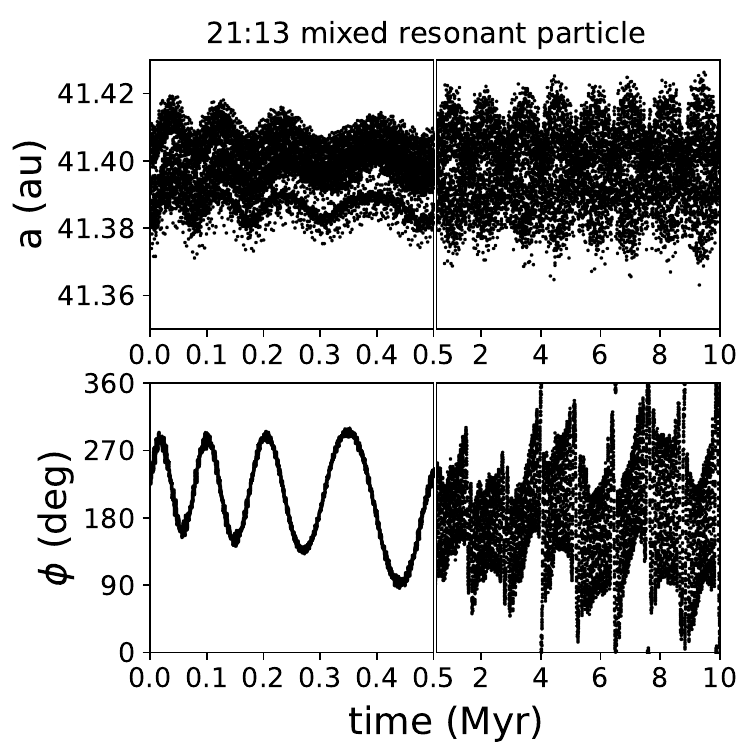}
    & \hspace{-15pt} \includegraphics[width=2in]{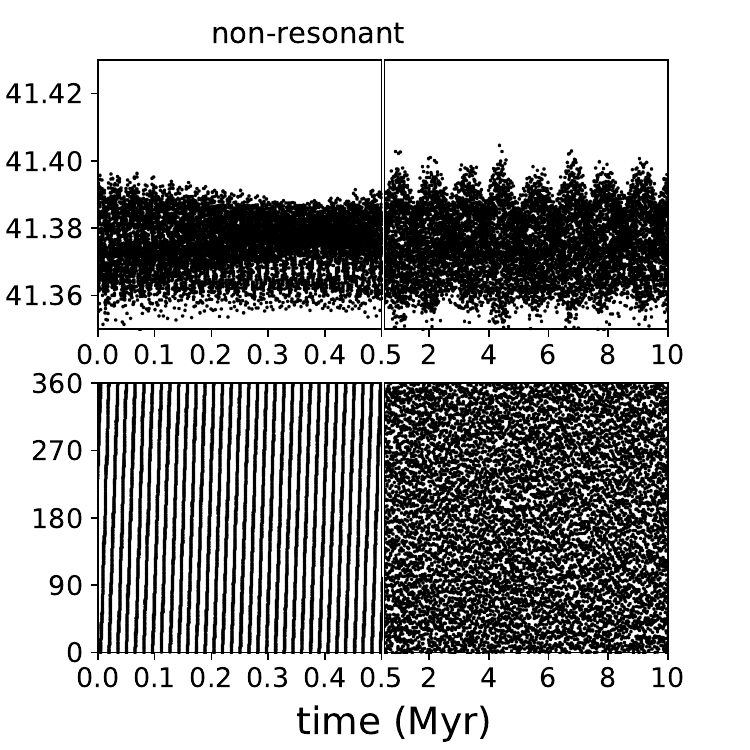} \\
    \end{tabular}
    \caption{Evolution of two clones of TNO 2014 UJ299 in semimajor axis (top panels) and resonant angle $\phi$ (bottom panels). 
    The left panels show a clone librating in a high-order mixed-$e$-$i$-type resonance with resonant angle $\phi = 21\lambda - 13\lambda_N - 4\varpi - 4\Omega$, whereas the right panels show a nearby non-resonant clone. 
    In each panel the time-axis is discontinuous with the left portion showing the high-resolution output from 0-0.5 Myr and the right portion showing lower-resolution output from  0.5-10 Myr. 
    While the high-resolution output does reveal some differences in the semimajor axis behavior between the two clones, they look nearly identical over 10 Myr timescales. 
    Their inclinations and eccentricities also evolve nearly identically. 
    So we are left to ponder about the dynamical significance of the libration of this very high-order resonant argument. }
    \label{f:high-order-res}
\end{figure}

\textcolor{black}{When classifying TNOs manually, visual inspection of the semimajor axis evolution by an experienced evaluator is essentially always sufficient to identify resonant behavior, with  inspection of the resonant angle used only to confirm the classification of the $a$ evolution and to identify the specific resonance. 
This, and the computational cost described above for calculating resonant angles, motivates us to discard resonant angles for our machine learning classifier.}
We will instead rely solely on characterizations of the standard set of orbital elements and position data, including calculating parameters designed to mimic the resonant characteristics and patterns that are  apparent by visual examination by a human (see Section~\ref{ss:features}).
However, this approach also has some downsides.
Sometimes an extensive resonant angle search reveals libration of very high-order resonant arguments that would not be readily apparent from visual inspection of the orbital evolution. 
Figure~\ref{f:high-order-res} shows the semimajor axis evolution from a short, high-resolution-output integration and a longer, lower-resolution-output integration for two slightly different orbits within the observational uncertainties of TNO 2014 UJ299. 
We find that the orbit shown in the left panels librates in the high-order mixed-$e$-$i$-type resonance characterized by $\phi = 21\lambda - 13\lambda_N - 4\varpi - 4\Omega$ while the orbit in the right panels shows no libration of any resonant angle that we tested.
Aside from the behavior of that resonant angle and the short-term semimajor axis evolution of the resonant clone which appears \textit{slightly} different (left panels),  the longer-timescale evolution is nearly indistinguishable between the two clones and the resonance has no effect on their $e$ or $i$ evolution.
\textcolor{black}{A comparison of periodograms in $a$ for these two particles does not provide substantially different information than this visual assessment of Figure~\ref{f:high-order-res}.}
As we will find in Section~\ref{ss:performance}, a significant fraction of the mis-classifications made by our machine learning classifier are failures to identify cases of very high-order ($q\ge10$) resonances such as the one in Figure~\ref{f:high-order-res}.
But perhaps the fact that these resonant orbits are so difficult to distinguish from nearby non-resonant orbits both by eye and by analyses of their orbital evolution hints at a limit to the true dynamical relevance of such high-order resonances.

Even after a resonance has been identified, one challenge remains: deciding how and where to draw the line between mostly resonant behavior and mostly non-resonant behavior. 
The \cite{Gladman:2008} scheme chooses to label cases that librate in $\phi$ for at least 50\% of a 10 Myr simulation as resonant and those librating less than 50\% as non-resonant.
But even that determination can be challenging for objects like the one shown in the left panels of Figure~\ref{f:messy-res}.
When trying to automate classification, whether the libration meets the 50\% threshold depends strongly on the window size over which the time series is examined, the time-sampling of $\phi$, the exact tolerances on the maximum range in $\phi$, etc. 
Visual examination of $\phi_{7:4}$ in Figure~\ref{f:messy-res} places this object in the resonant category, though even that is a bit of a judgement call as it is very close to the 50\% threshold; and as we will discuss in the next section, visual examination and human judgment is not a sustainable approach to this problem. 
We will show in Section~\ref{ss:performance} that many of the mis-classifications by the machine learning classifier involve objects that display intermittent resonant interactions.
It is unsurprising that these edge cases remain a challenge whether being classified by humans or via machine learning. 

\subsection{The need to improve automated classification}

The strongest motivation for automating TNO classification is the expected $\sim40,000$ new TNOs that will be discovered LSST \citep{Ivezic:2019,Schwamb:2019}. 
Even considering just best-fit orbits, manual classification of such a large number of TNOs would be daunting.
The \cite{Gladman:2008} scheme for determining whether a real TNO's dynamical classification is secure or insecure requires classifying three variants on each TNO's orbit (the best-fit orbit, and a minumum- and maximum-semimajor axis orbit), tripling the number of integrations to be visually inspected.
While the three-clone approach is quite useful, it does not yield probabilities of dynamical class membership for a TNO.
Additionally, for objects with large orbital uncertainties, the gaps between the best-fit and extreme clones can be too large to adequately sample the range of possible dynamical classes.
We ideally would like to run many more clones (100 or more) of each observed TNO orbit to fully sample the dynamical classes consistent with the observations and provide meaningful probabilities for the most likely dynamical class.
For the expected LSST TNO sample, this aspiration means classifying millions of orbits. 
At this scale, human assessment must be removed from the classification process.

Machine learning classifiers are well-suited for the task of replacing human inspection in the TNO classification process. 
\cite{Smullen:2020} described a first pass at training and testing a Gradient Boosting Classifier for assigning dynamical classes according to the \cite{Gladman:2008} scheme based on a simple set of data features in short integrations. 
This classifier achieved 97\% accuracy \textcolor{black}{(the percentage of all predictions that are correct)} on a set of TNOs that met the \cite{Gladman:2008} criteria for secure dynamical classifications. 
The accuracy dropped to 75\% when considering only insecurely classified TNOs, which is unsurprising as those are often insecurely classified because they exhibit a complicated mix of resonant and non-resonant behavior. 
As discussed above, these are the most challenging cases to classify.
When the \cite{Smullen:2020} classifier was applied to a new set of TNOs (with a typical mix of secure and insecure classifications as determined by the manual classification approach), the accuracy was 92.4\%. 
Given the relatively small training and testing set available, and the limited set of data features used for the classifier, this was a very promising result. 
However, that classifier would still result in thousands of mis-classified TNOs from the LSST discoveries. 
In the next section we describe an updated and improved machine learning classifier based on a training and testing set an order of magnitude larger than that in \cite{Smullen:2020} and employing an improved set of data features. 
This new classifier provides the same classification as a human $97-98\%$ of the time and dynamically relevant classifications $>99\%$ of the time (see Section~\ref{ss:performance}), making it a reasonable replacement for human classification in the LSST era.

\section{Building a Machine Learning Classifier for TNOs}\label{s:classifier}

Building on the success of \cite{Smullen:2020}, we will take a supervised learning approach to TNO classification. 
This requires constructing a sample of correctly labeled TNO orbits, deciding what data from those orbits will be provided to the classifier, and then training, testing, and optimizing the classifier. 
We describe our training/testing set of TNO orbits and the dynamical labels we assign to those orbits in Section~\ref{ss:training-set}.
Section~\ref{ss:features} describes how we turn the integrations of those TNO orbits into a discrete set of data features to use in the classifier.
Finally, we train and test a variety of classifiers in Section~\ref{ss:performance} based on those dynamical labels and data features.

\subsection{Building and labeling an adequately large and diverse training dataset}\label{ss:training-set}

We built an initial training and testing set based on the observed set of multi-opposition TNOs pulled from the Minor Planet Center as of December 2023.
These TNOs were integrated and classified based on the \cite{Gladman:2008} scheme, which involves integrating 3 cloned particles representing variations of an observed TNO's orbit. 
The details of this set of observed objects and their resulting dynamical classifications are given in \cite{Volk:2024}.
For the purposes of this chapter, the important details are that we have orbital time series data for 9477 test particles representing 3159 real, observed TNOs and that these time series data have been examined visually to confirm that the evolution of each test particle is accurately labeled with the following information: 
\begin{enumerate}
    \item the \cite{Gladman:2008} classification as resonant, scattering, classical, or detached 
    \item if resonant, whether the resonant argument is for the typical $e$-type resonance or if it is for a mixed $e$-$i$-type resonance
    \item if resonant, whether the test particle's resonant libration is intermittent or remains clearly bounded for the entire 10 Myr integration.
    \item if classical or detached, whether the test particle experiences significant resonant interactions even though it is dominantly non-resonant; if a particle experiences short periods ($<5$~Myr total) of resonant libration or clearly crosses back and forth across the separatrix of a resonance, it is labeled as `resonant-interacting'.
\end{enumerate}
The integration outputs include the typical orbital elements (barycentric $a$, $e$, $i$, $\Omega$, $\varpi$, argument of perihelion $\omega$, and mean anomaly $M$) for the particle and the giant planets.
We have high-resolution output for the first 0.5 Myr (outputs every 50 yr; 10,000 outputs per particle) and lower-resolution outputs over the 10 Myr timespan (outputs every 1000 yr; 10,000 outputs per particle).
Plots of both timescales were examined to assign the \cite{Gladman:2008} classifications and additional labels listed above.
Table~\ref{t:training-set} provides details on the number of particles in this dataset divided into the different dynamical categories.

\begin{table}[]
    \centering
    \begin{tabular}{c|c|c|c}
         & number of real  & number of synthetic   & total \\
           & TNO particles &  TNO particles  &  \\
       \hline
        resonant & 3203 & 4840 & 8043 \\
        scattering & 712 & 658 & 1370 \\
        detached & 618 & 372 & 990 \\
        classical & 4944 & 28 & 4972\\
        \hline
        $e$-type resonant & 3085 & 3762 & 6847\\
        mixed-$e$-$i$ resonant & 119 & 1078 & 1197\\ 
        \hline
        cleanly librating & 2877 & 3226 & 6103 \\
        intermittently librating & 327 & 1614  & 1941 \\ \hline
        $p$:$q$ resonant, $q\ge5$ & 336 & 3023 & 3359\\ 
        $p$:$q$ resonant, $q\ge10$ & 63  & 325 & 388 \\ \hline
        non resonance interacting &&&\\classical/detached & 4897 & 104 & 4793\\ \hline
    \end{tabular}
    \caption{Classifications of the 15375 particles in the training set. The top four rows show the number of test particles classified into each of the four \cite{Gladman:2008} dynamical classes. Below that are divisions by additional dynamical detail. For each category, we indicate how many of the test particles are generated directly from observed TNO orbits and how many are generated from synthetic TNO orbits.}
    \label{t:training-set}
\end{table}

Some of the dynamical categories are not well-sampled in the set of observed TNOs, notably high-order resonances ($p$:$q$ MMRs with $q\ge5$) and mixed-$e$-$i$ resonances.
Scattering and detached particles are also less numerous than classical and resonant ones; scattering, detached, and distant resonant TNOs spend most of their time very far from the Sun compared to classical and closer-in resonant TNOs and thus suffer from stronger observational biases (see, e.g., \citealt{Gladman:2021} for a discussion of observational biases in the TNO populations). 
To augment the training and testing set, we generated synthetic TNO orbits targeting these under-sampled populations (see Appendix~\ref{a:synthetic} for details of how these orbits were generated) and integrated and classified them just as above.
Table~\ref{t:training-set} lists the number of synthetic TNOs of each dynamical type that were added to the dataset. 
We increased the number of mixed-$e$-$i$ resonant particles and high-order resonant particles by an order of magnitude, more than doubling the total resonant training set.
We also nearly doubled the number of scattering particles and increased the number of detached particles by 60\%.
This enhanced training set still does not fully sample the range of orbits possible for TNOs, but it provides significantly more examples of orbits that are underrepresented in the observationally-based data; this ensures that there are enough of these rarer object types to both train and test the classifier. \textcolor{black}{As we will discuss in Section~\ref{ss:performance}, the inclusion of these synthetic TNOs significantly decreases the rate at which the classifier provides dynamically irrelevant classifications.}

The total testing and training dataset comprises integrations of 15,375 particles with human-assigned dynamical labels. 
Unlike the training set previously used in \cite{Smullen:2020}, this set is not limited to just the observed set of TNOs, and it is not limited to only TNOs with secure dynamical classifications. 
We have also labeled each particle with information not contained in the \cite{Gladman:2008} classifications that will enable us to test the ability to divide TNOs into different classes besides the four standard ones.
The next section describes the analyses performed on the integration outputs of our training and testing set orbits to prepare them for the classifier.

\subsection{Choosing appropriate and useful time series data features for the classifier}\label{ss:features}

The type of classifier we employ in Section~\ref{ss:performance} is not given a set of rules by which to classify TNOs, and it does not ingest the entire time series data of each particle it classifies. 
It is instead given a set of so-called data features, which are parameters calculated to summarize important aspects the time series data.
It uses a training set of data features that are labeled by class to construct a set of rules for how to use those data features to predict the correct class. 
The rules constructed from the training set are then tried out on a testing set of data features, and the predicted classes are compared to the known correct classes to determine the classifier's accuracy. 
See Chapter 1 for a more detailed description of this kind of machine learning classifier.

We use data features calculated from both the short, high-resolution orbital integration time series and the long, lower-resolution time series.
We begin by considering the same set of data features as in \cite{Smullen:2020}, which are:  1) the minimum, maximum, mean, and standard deviation of each of the particle's orbital elements; 2) the minimum, maximum, mean, and standard deviation of the rate of change of each orbital element from one simulation output to the next; and 3) the maximum range of each of the orbital elements and its rate of change.
In \cite{Smullen:2020}, the orbital elements considered for these parameters were $a$, $e$, $i$, $\Omega$, and $\omega$, and their values were all calculated from just a short $10^5$ year integration.
In addition to calculating these features from both short and long timescale integrations (to accommodate some of the additional data features we describe below), we also make some modifications to this set of simple features.
First, we add the longitude of perihelion ($\varpi=\Omega+\omega$), the perihelion distance ($q$), and the Tisserand parameter with respect to Neptune ($T_N = a_N/a + 2\sqrt{a/a_N(1-e^2)}\cos i$) to the list of time series variables considered.
We also discard most of the features based on the absolute values of the orbital angles $\Omega$ and $\varpi$ because the initial values of these angles are subject to strong observational biases based on where in the sky surveys have looked for TNOs (see, e.g., \citealt{Shankman:2017}).
Resonant objects are subject to additional, epoch-dependent observational biases in $\varpi$ \citep[e.g.][]{Gladman:2012,Volk:2020}.
Even though these biases might wash out in the longer integrations as these angles precess, and there is some important dynamical information in their overall distribution \citep[e.g.][]{JeongAhn:2014}, we do not wish to potentially bias the classifier.
We keep the features calculated based on the rate of change for each of these angles.
We also keep the min, max, mean, and standard deviation for $\omega$ because the dynamical information contained in the $\omega$ evolution outweighs the potential for introducing biases; many resonant TNOs undergo so-called Kozai libration of $\omega$ (see, e.g., \citealt{Gomes:2005}), meaning specific values for the average and range of $\omega$ could be an important marker for resonant behavior. 
For $a$ and $q$, we also include values of the standard deviation and maximum range normalized to their average values.

Next we consider a range of new data features.
Following \cite{Volk:2020}, we calculate a spectral fraction for the evolution of each particle's semimajor axis, components of the eccentricity and inclination vectors ($e\sin{\varpi}$ and $\sin i\sin\Omega$), and the angular momentum deficit (\textsc{amd}=$a(1-\sqrt{1-e^2})\cos i$).
The spectral fraction is a parameter that captures whether the evolution of a time series parameter is dominated by just a few frequencies or contains many, potentially overlapping frequencies.
We define the spectral fraction of a time-series variable by taking a Fast Fourier Transform (FFT) of that time-series and determining what fraction of the frequencies in that FFT have an associated power (defined as the amplitude squared at that frequency) greater than 5\% of the highest-amplitude frequency's power.
A small spectral fraction means the single, highest-amplitude frequency dominates the evolution while a large spectral fraction means there are many frequencies affecting the evolution.
\cite{Volk:2020} found that for simulated multi-planet systems, the spectral fraction could be used as an indicator of long-term stability or instability.
It is likely that the same holds true for TNO orbits as those with many overlapping frequencies are more likely to experience chaotic evolution.
Additionally, the semimajor axis evolution of resonant and non-resonant TNOs are very different in the frequency domain; the resonant libration of $a$ adds a powerful, lower-frequency term to the $a$ evolution of resonant objects. 
For $a$ and the \textsc{amd}, we calculate a spectral fraction from both the short and long data series because both simulations cover timescales relevant to the dynamics.
The dominant timescales for the eccentricity and inclination evolution are typically on Myr timescales, so we only calculate the spectral fraction for those parameters from the 10 Myr time series.
For each parameter, we also include as data features the power of the highest-amplitude FFT frequency, the summed power of the top three FFT frequencies, and the values of the top three FFT frequencies.

\begin{figure}
    \centering
    \includegraphics[width=2.7in]{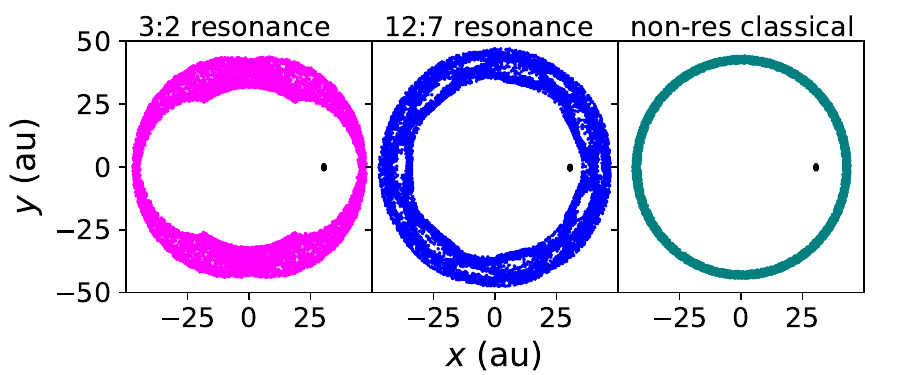}
    \includegraphics[width=2.7in]{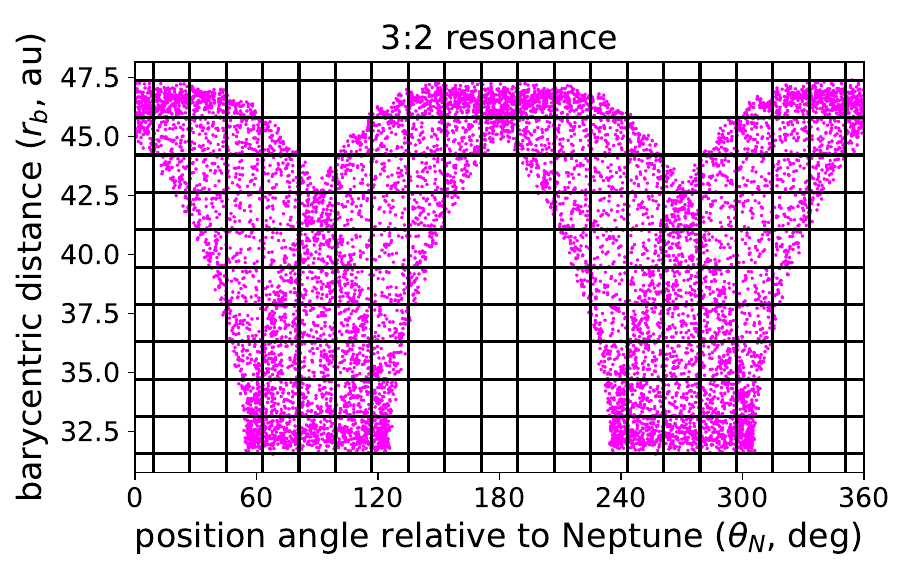}
    \includegraphics[width=2.7in]{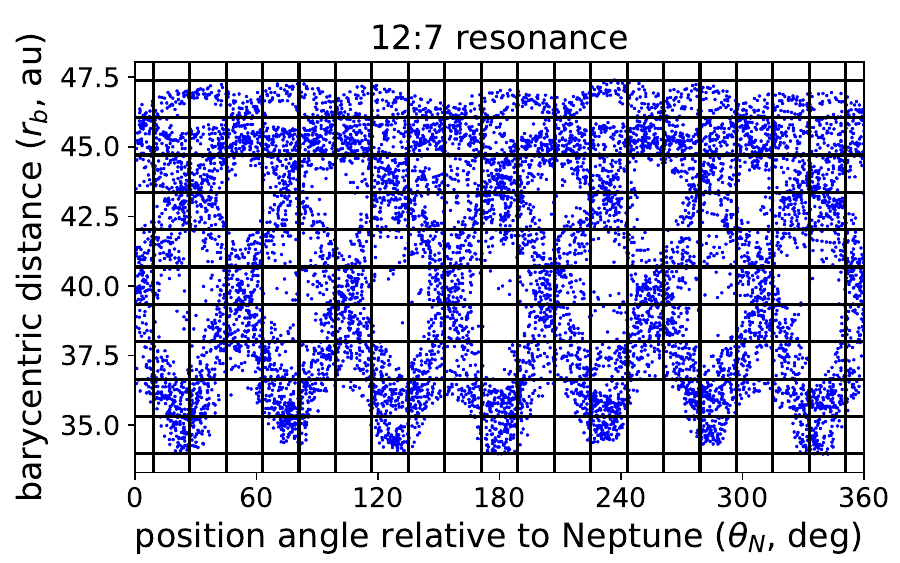}
    \includegraphics[width=2.7in]{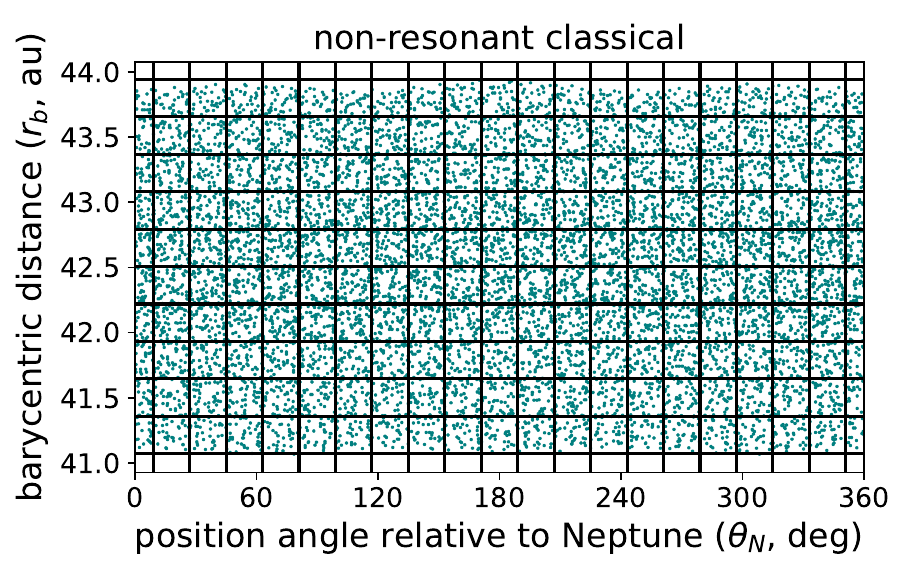}
    \caption{Top panel: Position of three TNO orbits (colored) and Neptune (black) over 10 Myr in a frame rotating at Neptune's instantaneous azimuthal rate. 
    Bottom three panels: Barycentric distance, $r_b$ vs longitude angle relative to Neptune, $\theta_N$, at every output over the 10 Myr integrations for the three TNOs in the top panel. 
    Neptune would be centered at the point (0,30.06) in these plots.
    We divide the evolution of the particle in this plane into a 10 by 20 grid, with the 10 bins in $r_b$  bounding the particle's minimum and maximum barycentric distances; the grid in $\theta_N$  starts and ends with half a bin so that Neptune is centered in the wrapped bin. 
    We then calculate data features based on this grid, including: the number of empty grid spaces overall as well as in the smallest distance range (near perihelion) and the largest distance range (near aphelion); the average and standard deviation in the number of points in all the grid spaces as well as in the perihelion and aphelion grid spaces. 
    }
    \label{fig:rotating-frame}
\end{figure}

We also consider data features related to the spatial distribution of a particle's path in a frame rotating at Neptune's instantaneous azimuthal rate.
Particles in resonance with Neptune follow distinct paths in this rotating frame as they come to their resonant perihelion locations relative to Neptune (see, e.g., \citealt{Malhotra:1996,Gladman:2012,Gladman:2021}).
For each particle, we transform the simulation outputs to this rotating frame with Neptune and consider the distribution of the barycentric distance $r_b$ and the projected angle with respect to Neptune in the x-y plane, an angle we will denote as $\theta_N$. 
We then parameterize the distribution of points in this reference frame by dividing a particle's range of barycentric distances into 10 bins, and $\theta_N$ into 20 bins from 0-360$^\circ$, yielding a 200 space grid.
Figure~\ref{fig:rotating-frame} illustrates the distribution of $r_b$ and $\theta_N$ across this grid for two resonant orbits and one classical belt orbit. 
For the distance bin that includes the particle's perihelion, we calculate a Rayleigh parameter \citep[e.g.][]{Fisher:1993} to describe how uniformly $\theta_N$ is distributed. 
This parameter is given by
\begin{equation}
    R = \sqrt{(<\sin\theta_N>)^2 + (<\cos\theta_N>)^2}
\end{equation}
where the averages are taken over all values of $\sin\theta_N$ and $\cos\theta_N$ with barycentric distances in the closest $r_b$ bin.
A particle with a uniform distribution in $r_b$ and $\theta_N$ would have $R=0$ while one that was perfectly concentrated at a single value of $\theta_N$ would have $R=1$. 
For the classical belt object in Figure~\ref{fig:rotating-frame}, $R=0.008$.
Particles in $p$:$q$ resonances where $q\ne1$ also have small values of $R$ because they come to perihelion at more than one $\theta_N$ value; the 3:2 resonant particle in Figure~\ref{fig:rotating-frame} has $R=0.007$. 
So we must generalize our calculation of $R$ for $p$:$q$ resonances to:
\begin{equation}
    R_{q} = \sqrt{(<\sin q \theta_N>)^2 + (<\cos q \theta_N>)^2} \label{eq:rz}
\end{equation}
for $q=1,2,3,..,10$; we do not consider $R_q$ for higher-order resonances as the values become less statistically distinguishable from uniform in this parameter with additional perihelion locations.
We take the maximum value of $R_{1-10}$ as the data feature for the classifier ($R_{max,peri}$).
For our example 3:2 particle, $R_{max,peri} = R_2 = 0.65$; similarly, our example 12:7 particle in Figure~\ref{fig:rotating-frame} has $R_{max,peri} = R_7 = 0.80$.
For every particle, we also calculate an analogous feature using the maximum $r_b$ bin, $R_{max,apo}$.

We calculate several features based on the grid in $r_b$ and $\theta_N$. 
We determine the number of empty grid spaces across the minimum and maximum $r_b$ bins ($n_{empty,peri}$ and $n_{empty,apo}$) and the standard deviation in the number of visits per grid space in these bins ($\sigma_{n,peri}$ and $\sigma_{n,apo}$); all of these parameters will be larger for resonant particles than non-resonant ones.
For the minimum $r_b$ bins, we determine how many of the bins on either side of Neptune's location at $\theta_N=0$ are empty ($n_{empty,N}$); this will be highest for low-order resonances, smaller for high-order resonances, and close to zero for non-resonant particles.
We also determine the standard deviation in the number of visits to each grid space across the entire 200 grid spaces ($\sigma_n$), the number of empty grid spaces ($n_{empty}$), the average and standard deviation in the visits to each non-empty grid space ($\overline{n}_{nz,avg}$,$\sigma_{n,nz}$), and the difference between the overall and non-zero standard deviations ($\Delta\sigma_n = \sigma_{n,nz} - \sigma_{n}$).
These data features are calculated for both the short and long simulations.

\begin{figure}
    \centering
    \includegraphics[width=3in]{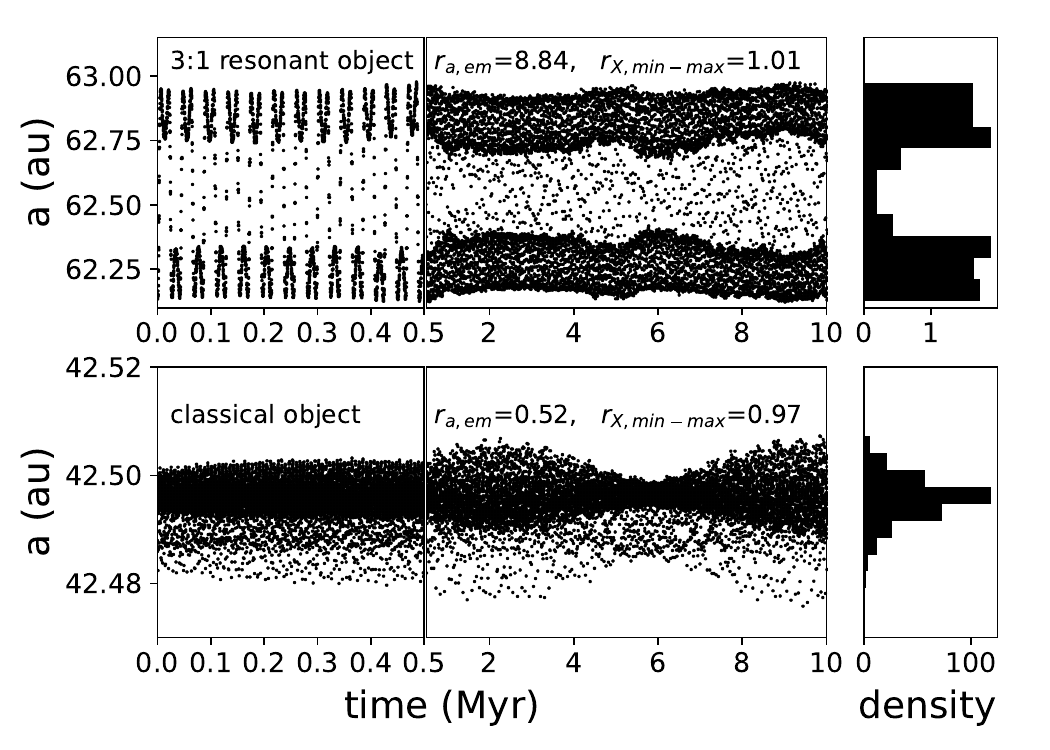}
    \caption{Semimajor axis time-series for a 3:1 resonant object (top panels) and a non-resonant classical object (bottom panels). In the left panels, the time-axis is discontinuous with the first portion showing the high-resolution output over 0-0.5 Myr and the remaining portion showing lower-resolution output over  0.5-10 Myr. 
    The right panels show the histogram of $a$ values across the 10 Myr integration, highlighting the difference between resonant and non-resonant evolution.
    In the machine learning classifier, the two data features,$r_{a,min-max}$ and $r_{a,em}$ related to the semimajor axis (see Table~\ref{t:features}), encapsulate the information in these histograms. }
    \label{fig:a-bins}
\end{figure}

We consider a few additional data features describing the $a$, $e$, and $i$ time-series.
Resonant TNOs often experience large amplitude quasi-periodic variations in their semimajor axes as a result of the resonant perturbations, and they spend more time at the extreme ends of their $a$-range than at the middle.
In contrast, completely non-resonant particles tend to have more uniform variations in $a$ within a smaller range; this is illustrated in Figure~\ref{fig:a-bins}. 
Particles interacting with a resonance but not librating can often have a lop-sided distribution in $a$ where they spend most of their time at one extreme compared to either the middle of the other extreme.
We chose a simple binning scheme in $a$ to help parameterize these three kinds of behavior: a bin near the minimum $a$ value spanning $a_{min}$ to $a_{min}+0.75\sigma_a$,  a bin near the mean $a$ value spanning $\overline{a}\pm0.375\sigma_a$, and a bin near the maximum $a$ value spanning $a_{max}-0.75\sigma_a$ to $a_{max}$. 
We then calculate the ratio of output $a$ values in the low-$a$ bin to those in the high-$a$ bin ($r_{a,min-max}$) as well as the average of the low- and high-$a$ bins to the average-$a$ bin ($r_{a,em}$).
The exact bin boundaries were chosen empirically based on comparisons of the ratios for typical detached, resonant, and classical TNOs.
We note that for scattering TNOs, those chosen bin boundaries sometimes overlap because their evolution is not well characterized by averages and standard deviations; in those cases we instead define the bins as the lower, middle, and upper 25\% of the full $\Delta a$ range.
In an attempt to parameterize intermittent resonant behavior, we also examine the maximum change in $r_{a,min-max}$ and $r_{a,em}$ calculated over four non-overlapping time ranges in both the short and long integrations.
For $e$ and $i$, we apply a similar scheme as above to characterize their distributions in both the short and long integrations, though we do not examine the time-dependency of these ratios; these ratios for $e$ and $i$ do not turn out to be as useful to the classifier as their corresponding ratios for $a$, but we include them for completeness.
Finally, we calculate correlation coefficients between the $a$ and $e$, $a$ and $i$, and $e$ and $i$ simulation outputs ($C_{ae}$, $C_{ai}$, and $C_{ei}$) to capture whether the variations in these elements are coupled.

\begin{table}
    \centering
    \begin{tabular}{l|l}
    feature name & brief description\\ \hline \hline
     & basic time-series data features \\ \hline
     $\overline{X}$    &  average of $X$ where $X=a,e,i,\omega,q,T_N$\\
     $\sigma_{X}$    &  standard deviation in $X$ for $X=a,e,i,\omega,q,T_N$\\
     $\sigma_{X,normed}$    &  $\sigma_{X}/\overline{X}$  for $X=a,q$\\
     $X_{min}$    &  minimum value of  $X$ for $X=e,i,\omega,q,T_N$\\
     $X_{max}$    &  maximum value of  $X$ for $X=e,i,\omega,q,T_N$\\
     $\Delta X$   & $X_{max}$ -$X_{min}$ for $X=a,e,i,\omega,q,T_N$\\ 
     $\Delta X_{normed}$   & $\Delta X/\overline{X}$ for $X=a,q$\\
     $\overline{\dot{X}}$  &  average $dX/dt$  \\ 
                      & for $X=a,e,i,\omega,\Omega,\varpi,q$\\
     $\dot{X}_{min}$  &  minimum value of $\dot{X}$ for $X=a,e,i,\omega,\Omega,\varpi,q$\\
     $\dot{X}_{max}$  &  maximum value of $\dot{X}$ for $X=a,e,i,\omega,\Omega,\varpi,q$\\
    $\sigma_{\dot{X}}$ & standard deviation in ${\dot{X}}$ for $X=a,e,i,\omega,\Omega,\varpi,q$ \\
    $\sigma_{\dot{X}, normed}$ & $\sigma_{\dot{X}}/\overline{\dot{X}}$ for $X=\omega,\Omega,\varpi$ \\
    
    $\Delta\dot{X}$  &  $\dot{X}_{max} - \dot{X}_{min}$ for $X=a,e,i,\omega,\Omega,\varpi,q$ \\    
    $\Delta\dot{X}_{normed}$  & $\Delta\dot{X}/\overline{\dot{X}}$ for $X=\omega,\Omega,\varpi$ \\ \hline \hline
         & rotating frame data features based on binning barycentric distance $r_b$ and   \\
         & position angle relative to Neptune $\theta_N$; `near perihelion' and `near aphelion' \\
         & indicate the minimum and maximum $r_b$ bins\\ \hline
    $n_{empty,peri}$  & number of empty bins in $\theta_N$ near perihelion  \\
    $n_{empty,N}$ & number of empty bins surrounding Neptune in $\theta_N$ near perihelion\\
    $\sigma_{n,peri}$ & standard deviation in the number of visits per $\theta_N$ bin near perihelion\\
    $\Delta{n}_{peri}$ & maximum difference in number of visits between $\theta_N$ bins near perihelion \\
    $n_{empty,apo}$  &  number of empty bins in $\theta_N$ near aphelion  \\
    $\sigma_{n,apo}$ & standard deviation in the number of visits per $\theta_N$ bin near aphelion\\
    $\Delta{n}_{apo}$ & maximum difference in number of visits between $\theta_N$ bins near aphelion \\
    $R_{max,peri}$ & maximum of Eq.~\ref{eq:rz} for $q=1$ through $q=10$ near perihelion\\
    $R_{max,apo}$ & maximum of Eq.~\ref{eq:rz} for $q=1$ through $q=10$ near aphelion\\
    $n_{empty}$ & number of empty bins in the $r_b$-$\theta_N$ grid \\
    $\sigma_n$ & standard deviation in the number of visits across all bins\\
    $\overline{n}_{nz}$ & average number of visits per bin across all non-empty bins\\
    $\sigma_{n,nz}$ & standard deviation in the number of visits across all non-empty bins\\
    $\Delta\sigma_n$ & $\sigma_{n,nz} - \sigma_{n}$\\
    \hline \hline
                & FFT features for $X=a,(e\sin\varpi),(\sin i \sin\Omega),$\textsc{amd} \\ \hline
    $X_{sf}$      & spectral fraction of $X$ \\ 
    $f_{X,i=1,2,3}$& The three most powerful frequencies from an FFT of parameter X \\
    $P_{X,max}$ & Power associated with the peak frequency from an FFT of parameter X \\
    $P_{X,max3}$ & Sum of the power associated with the 3 most powerful frequencies from an \\
                &  FFT of parameter $X$ \\               
    \hline \hline
                & other time-evolution features \\ \hline
    $r_{X,em}$ & ratio of time spent near $X_{min}$ and $X_{max}$ to $\overline{X}$ for $X=a,e,i$\\
    $r_{X,min-max}$ & ratio of time spent near $X_{min}$ to $X_{max}$ for $X=a,e,i$\\    
    $\Delta r_{a,em}$ & maximum change in $r_{a,em}$ over 4 time windows in the simulation \\
    $\Delta r_{a,min-max}$ & maximum change in $r_{a,min-max}$ over 4 time windows in the simulation \\
    $C_{ae}$ & correlation coefficient between $a$ and $e$\\
    $C_{ai}$ & correlation coefficient between $a$ and $i$\\
    $C_{ei}$ & correlation coefficient between $e$ and $i$\\

    \end{tabular}
    \caption{List of the data features calculated from the integrations.}
    \label{t:features}
\end{table}

Table~\ref{t:features} lists all of the data features we calculate to provide to the classifier. 
In total, we have 227 data features from the short and long integrations.

\subsection{Performance of a Gradient Boosting Classifier for TNO classification}\label{ss:performance}

Here we train and test a classifier using the data features described above with different sets of class labels.
We chose to use the scikit-learn \citep{scikit-learn} GradientBoostingClassifier\footnote{\url{https://scikit-learn.org/stable/modules/generated/sklearn.ensemble.GradientBoostingClassifier.html}}.
\cite{Smullen:2020} found that this was the best-performing classifier within scikit-learn, and we confirmed that by testing the other classifiers within sklearn.ensemble\footnote{\url{https://scikit-learn.org/stable/modules/classes.html#module-sklearn.ensemble}} that support multi-label datasets.
For all tests described below, we set the following parameters, which were found to optimize the performance of GradientBoostingClassifier for the standard set of \cite{Gladman:2008} classes \textcolor{black}{based on an initial round of training and testing using just the real TNOs and an initial subset of our final data features}:
max\_leaf\_nodes = None, min\_impurity\_decrease = 0.0, min\_weight\_fraction\_leaf = 0.0, 
min\_samples\_leaf = 1, min\_samples\_split = 3, 
criterion = `friedman\_mse', subsample = 0.9, learning\_rate = 0.15, max\_depth=8, max\_features = `log2', n\_estimators = 300.
\textcolor{black}{We found this set of parameter values by starting with the default values and making changes to one parameter at a time in moderate increments away from the default (increments of 0.05 for float parameters ranging from 0-1, increments of 1 for most integer parameters, and increments of 50 for n\_estimators), stopping at a local maximum in the classifier's accuracy. 
A repeat of this optimization process using the final set of data features and our entire training and testing set yields identical results for all parameters except min\_samples\_split. 
We find that increasing min\_samples\_split to 4 for the final classifier would yield a 0.05\% improvement in performance; we judged this improvement too small to include in the analysis below.}
For each version of the classifier described below, we use the dataset described in Section~\ref{ss:training-set} with all of the data features listed in Table~\ref{t:features}, training the classifier on 67\% of the dataset (10,156 particles) and testing on the remaining 33\% (5071 particles).
We held the division of the training and testing sets fixed when testing the accuracy of the classifier in predicting different sets of dynamical classes.

We begin by training and testing the classifier based on the \cite{Gladman:2008} classes described in Section~\ref{s:classes}: resonant, classical, scattering, and detached.
\textcolor{black}{We discuss the performance of the classifier in terms of its accuracy, defined as the percentage of all predictions that are correct.
We note that for this classifier, other common performance metrics such as precision and recall yield nearly identical percentages (to within $\sim0.02$\%), so we do not list them.}
The accuracy of the classifier in assigning the same classes as the human classifier is 97.3\%. 
If we consider the subset of the classifier's predictions that are given at $>99\%$ confidence (4853 of the 5071 predictions), the accuracy increases to 98.7\%. 
Given the fuzzy nature of some of the dynamical class boundaries (see discussion in Sections~\ref{s:classes} and~\ref{ss:resonances}), it is useful to divide the misclassified particles into four categories: 
\begin{itemize}
    \item not actually incorrect classifications -- These are cases where, upon further inspection, the classifier was correct and the human-assigned class was incorrect (almost always high-order resonances that were missed during visual inspection).
    \item trivial misclassifications -- These are cases where the classifier places a particle on the wrong side of either the classical/detached boundary or the scattering/detached boundary. As discussed in Section~\ref{s:classes}, the eccentricity boundary between classical and detached is based more on cosmogonic arguments than on present-day dynamics, and the classifier is not aware of the exact $e$ cut.  Similarly, the $\Delta a=1.5$~au threshold is a somewhat arbitrary and the classifier is not given this rule. Both of these types misclassifications could be trivially swapped during a post-processing check of the classifier's predictions. (Note that some of the swapped classifications would still be incorrect if, for example, the particle is truly resonant.)
    \item marginally wrong classifications -- These are cases where the assigned class describes \textit{part} of the particle's evolution, but doesn't match the human-classifier's assessment of the majority of the evolution. Examples include: labeling a particle resonant when it librates some of the time, but does not meet the 50\% threshold; labeling an intermittently resonant particle that only just meets the 50\% threshold as classical or detached.
    \item completely incorrect classifications -- These are cases where the class predicted for the particle is unrelated to its evolution.
\end{itemize}

For the \cite{Gladman:2008} dynamical classes, the classifier made predictions that differ from the human-assigned classes for 135 particles out of 5071. 
Of these, 15 particles were trivially mis-classified (5 at the classical/detached boundary, 10 at the scattering/detached boundary) and there were 4 cases where the human-assigned labels were wrong (human error!) and the classifier was correct. 
This brings the classifier's accuracy up to 97.7\%.
Of the remaining 116 incorrectly classified particles, 50 were predicted to be classical or detached when they were actually resonant. 
Most of those particles (42 out of 50) display only intermittent resonant libration, so they fall into the marginally wrong category in that they do display significant non-resonant behavior.
Only 8 cleanly librating resonant particles were fully incorrectly classified as non-resonant; of these, 7 are in very high-order resonances ($q\ge10$; including 3 mixed-$e$-$i$ type) in the classical belt region and one is in the distant 54:5 resonance with Neptune.
The classifier misclassified 51 classical or detached particles and 1 scattering particle as resonant. 
Of these 52 particles, 44 show intermittent resonant libration, so only 8 are fully incorrectly classified as resonant while displaying no resonant behavior.
Finally, the classifier incorrectly classified 10 resonant particles and one detached particle as scattering.
Eight of the resonant particles librate in resonances for more than half the simulation (and thus are `correctly' labeled as resonant) before weakly scattering out of resonance; so the classifier is only marginally wrong as they do scatter.
The classifier was only entirely wrong about scattering behavior in 3 cases; all 3 cases involve particles librating in or interacting with the wide, symmetric libration zones of N:1 resonances \citep[see, e.g.,][]{Lan:2019}, which can mimic the large semimajor axis variations associated with the scattering population.

Overall, the classifier only assigned completely irrelevant labels in 19 cases, or 0.4\% of the time; an additional 1.9\% of the predictions were only marginally incorrect.
Nearly half (7) of the fully incorrectly labeled particles are in very high order ($q\ge10$), weak resonances in the classical belt region discussed in Section~\ref{ss:resonances}. 
The orbital evolution of these particles are so weakly affected by the libration in those resonances that it is unsurprising that the classifier did not identify them. 
When we restrict ourselves to the subset of classifier predictions made at $>99\%$ confidence, only 0.14\% of the predictions were completely incorrect with an additional 1.1\% of the predictions being marginally incorrect.
The classifier performs remarkably well at identifying relevant dynamical behavior.

Given that the boundary between the classical and detached population is somewhat arbitrary and that the common feature between the two populations is non-resonant, non-scattering behavior, we also tested a simplified version of the \cite{Gladman:2008} scheme that combines them, giving us the classes: resonant, scattering, and classical/detached.
With this change, the classifier matched the human classifications in 97.7\% of all predictions and 98.8\% of the subset of predictions made at $>99$\% confidence (4876 out of 5071 particles).
This represents a small improvement in both the accuracy and the number of high-confidence predictions
Of the 113 incorrectly classified particles, 11 are trivially misclassified at the scattering/detached boundary, and one was incorrectly classified by the human, bringing the overall accuracy of the classifier up to 98\%.
Of the remaining 101 misclassifications, 50 are resonant particles that were incorrectly predicted to be classical/detached.
Only 6 of these particles experience clean libration for the entire 10 Myr timespan and are thus fully incorrectly labeled; all 6 are in high order resonances with $q\ge10$, including 2 mixed-$e$-$i$ resonant particles.
11 resonant particles were incorrectly predicted to be scattering; 9 of these are only marginal misclassifications because the particles do weakly scatter after spending most of the simulation in resonance.
There were 36 classical/detached particles incorrectly predicted to be resonant; all but 5 of these particles experience intermittent resonant libration and are thus only marginally misclassified.
Overall, the classifier provided only 13 completely incorrect classifications for this simplified classification scheme, an error rate of 0.25\%; an additional 1.7\% of the predictions were marginally incorrect.
Of the predictions made at $>99$\% confidence, only 6 are completely incorrect (half of which are in $q\ge10$ resonances), an error rate of 0.1\% with marginally incorrect predictions an additional 0.9\% of the time. 
While it is plausible that there are dynamical differences between the lower-$e$ non-resonant orbits in the classical belt region and the higher-$e$ non-resonant orbits of the more distant detached population, it seems slightly advantageous to combine them in the classification process (especially as they can be trivially separated after machine learning classification).

\textcolor{black}{The above results are based on our entire training and testing dataset, which includes synthetic TNOs generated to provide more examples of under-populated observational classes. 
If we exclude these synthetic TNOs from the training and testing set, the classifier doesn't perform as well.
While the classifier still matches the human-assigned classes for the simplified \cite{Gladman:2008} scheme 97.5\% of the time (compared to 97.7\% of the time with the full dataset), it assigns completely incorrect classifications at a higher rate; 1.1\% of predictions are dynamically irrelevant compared to only 0.25\% above.
A significant number of the completely incorrect classifications when the synthetic TNOs are excluded are cases where cleanly librating high-order or mixed-$e$-$i$-type resonant TNOs are incorrectly predicted to be classical/detached. 
Including more examples of these orbits in the training and testing set clearly improves the classifier's ability to identify them as resonant.}

\begin{figure}
    \centering
    \includegraphics[width=4.5in]{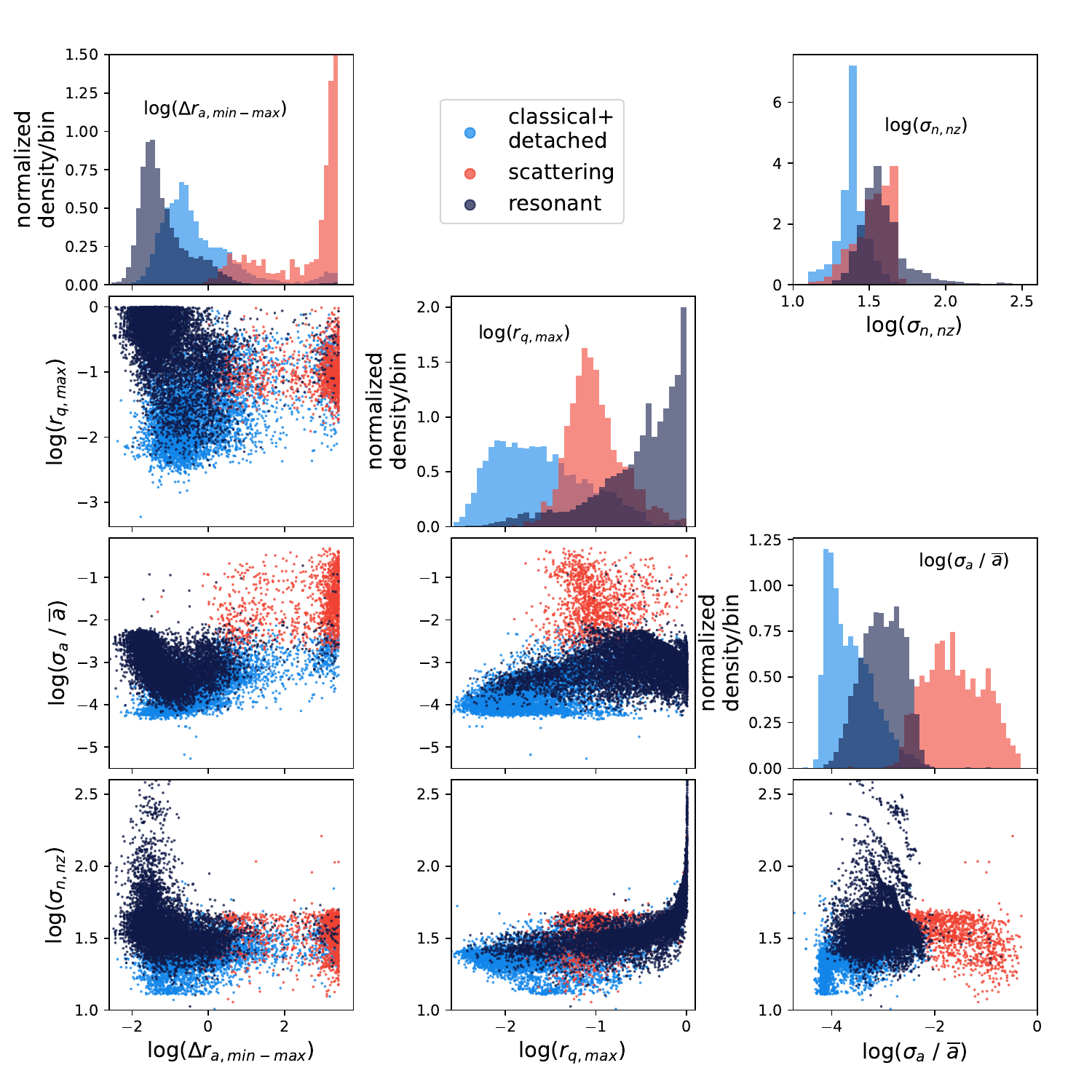}
    \caption{Scatter plots and histograms for four of the consistently highly-ranked data features across our entire training and testing dataset for resonant, scattering, and classical/detached TNOs.}
    \label{fig:features}
\end{figure}

We can examine which of the data features described in Section~\ref{ss:features} were relied upon the most by the classifier. 
We note that the exact ranking of the 227 features can vary significantly depending not just on which set of the above classifications were used, but also on the random seed used to divide the training and testing set or to initialize the classifier itself; there is clearly significant stochasticity to how the classifier uses the data features.
Thus we do not provide a full ranking list of the features, but instead note a subset of the features that tended to fall toward the bottom or top of the rankings.
Figure~\ref{fig:features} shows the distribution of four key features for the simplified \cite{Gladman:2008} classification scheme; shown are two features describing the semimajor axis evolution and two describing the particle's distribution in the rotating frame with Neptune.
Features that consistently were ranked highly by the classifier include most of the features describing the particle distribution in the rotating frame, most of the data features based on $a$ and many based on $\dot{a}$, some features based on $e$ and $\dot{\varpi}$, all the data features based on the FFT analysis of $a$ (peak frequencies and spectral fraction), and the $a-e$ and $a-i$ correlation coefficients.
Features that consistently ranked very low included all features based on $\dot{e}$ and $\dot{\Omega}$, the features based on FFT analysis of $e$, $i$, and \textsc{amd}, and the features based on $\omega$ (though those based on $\dot{\omega}$ fell in the middle of the rankings).
The 10 Myr maximum timescale of the analyzed time series is likely responsible for the less useful nature of the FFT analysis of $e$ and $i$; as discussed in Section~\ref{s:classes}, the secular timescales for the larger-$a$ TNOs are too long for the frequencies to be captured in a 10 Myr integration.
Similarly, 10 Myr is not always long enough to capture the $\omega$ libration that occurs inside some of Neptune's resonances, possibly explaining the relative unimportance of those data features.
The heavy emphasis on semimajor-axis based data features and those derived from the rotating frame is unsurprising given how strongly diagnostic those can be for resonant behavior.

\textcolor{black}{
For some of the consistently highly ranked data features above, we ran tests of the classifier with those features removed to see how the accuracy changes for the simplified \cite{Gladman:2008} scheme.
If we exclude all the simple time-series data features based on $a$ and $\dot{a}$, the classifier's accuracy drops slightly to 97.3\% (compared to 97.7\% with those features included), but the rate of dynamically irrelevant classifications rises to 0.9\% (compared to 0.25\%). 
Removing the FFT data features for $a$ reduces the accuracy to 96.9\% and increases the rate of dynamically irrelevant classifications to 0.9\%. 
Removing both sets of $a$ features reduces accuracy to 96.6\% while maintaining 0.9\% dynamically irrelevant classifications.
Removing the features calculated for the rotating frame resulted in an accuracy of 97.4\%, and increased the rate of dynamically irrelevant classifications to 0.7\%.
As we saw with the inclusion/exclusion of the synthetic TNO training and testing data, the rate of dynamically irrelevant predictions can be more sensitive to the selection of data features than the overall accuracy. 
This can also be the case when swapping two sets of features that are dynamically equivalent but different in scale (see recent discussion in \citealt{Smirnov:2024}). 
If we use mean motion in place of $a$ for the simple time-series features, the classifier's accuracy drops slightly to 97.5\% and the rate of dynamically irrelevant classifications rises to 0.5\%; including mean motion features in addition to $a$ features actually slightly further decreases the accuracy to 97.4\% though does not further increase the rate of irrelevant classifications.
We performed further testing of our classifier with different groups of features excluded (the broad categories of features separated in Table~\ref{t:features} as well as the basic time-series features for each individual element) and with each individual feature excluded. 
While some exclusions did not significantly decrease the accuracy of the classifier, we also did not find any improvements.
We conclude that our classifier performs best when all 227 of our data features are included. 
}

While the classifiers described above performed very well, we tested a few additional classification schemes. 
Starting with the simplified \cite{Gladman:2008} scheme above, we split the resonant TNOs into two classes based on $e$-type and mixed-$e$-$i$-type resonances, yielding four classes: scattering, $e$-resonant, $e$-$i$-resonant, and classical/detached.
After accounting for trivial misclassifications, this classifier predicted the same classes as the human 96.8\% of the time (98.4\% of the time for the 4835 predictions made at $>99\%$ confidence). 
Of the 159 misclassified particles, 37 were particles assigned to a different resonant class than assigned by the human classifier. 
However, 33 of these 37 actually show libration (often intermittent) of both the $e$- and $e$-$i$-type resonant arguments for their $p$:$q$ resonance; so these are not really incorrect classifications on the part of the machine learning or the human classifier as either label could be accurately applied.
Overall, the machine learning classifier only predicted completely incorrect classifications 0.4\% of the time, with marginally incorrect classifications another 2\% of the time.
For the $>99\%$ confidence predictions, only 0.15\% were completely incorrect (half of these are particles in $q>10$ resonances) and 1\% were marginally incorrect.
This performance is very similar to the simplified scheme without two resonant classes, though the utility of adding the $e$-$i$-resonant class is less clear given that many of the $e$-$i$-resonant TNOs also show libration in the $e$-type resonance.

We also tested a scheme to better separate the fully non-resonant classical/detached objects from their resonant-interacting counterparts. 
In this scheme we have the standard single resonant class, the scattering class, and then split the combined classical/detached population into classes of non-resonant and resonant-interacting.
The accuracy of this classifier was 95.7\% (97.9\% for the 4797 predictions made at $>99\%$ confidence), the lowest amongst those tested.
It also had the highest rate of completely incorrect predictions at 2.3\% of all predictions and 1.2\% of the high-confidence predictions.

For completeness, we also tried a very simple classification scheme, wherein we combined any particle in or interacting with a resonance into one class, anything stably non-resonant into a `non-res' class, and everything else as scattering. 
This classifier predicted the same classes as the human 97.2\% of the time (98.3\% of the time for the 4921 $>99\%$ confidence predictions).
But as in the scheme above, this scheme resulted in higher rates of completely incorrect classifications (1.5\% of the time across all predictions and 0.8\% of the time for the high-confidence predictions).
We conclude that using either the full or simplified \cite{Gladman:2008} dynamical classes results in the best-performing machine learning classifier.
\textcolor{black}{We provide an example jupyter notebook demonstrating the simplified \cite{Gladman:2008} classifier in the Github repository for this book.}

\section{Looking Forward to Future Applications and Improvements}\label{s:conclusions}

We have shown that machine learning can provide very accurate TNO classifications when there is a sufficiently large and diverse training set and when we provide the classifier with data features tailored to help identify resonant dynamics.
Our best classifier above, using a simplified version of the \citealt{Gladman:2008} dynamical classes to divide TNOs into resonant, scattering, and classical/detached TNOs, returned correct classifications 98\% of the time and dynamically relevant classifications (i.e., the particle displayed properties of the assigned dynamical class for at least part of its evolution) 99.7\% of the time. 
The classifier made high-confidence predictions 96\% of the time, with 99\% of these high-confidence predictions being correct and 99.9\% of them being dynamically relevant.

For the expected 40,000 TNOs from LSST, our classifier would only disagree with a human classifier for $\sim800$ TNOs, and only 100 of those would be assigned completely irrelevant dynamical classes.
For the expected $\sim38,400$ TNOs classified at $>99\%$ confidence, these numbers would drop to $\sim400$ and $\sim40$, respectively; and the number of lower-confidence predictions would be $\sim1600$, a number small enough for manual classification.
These are promising results that show machine learning is a viable replacement for manual classification for the LSST era.

\begin{figure}
    \centering
    \includegraphics[width=3in]{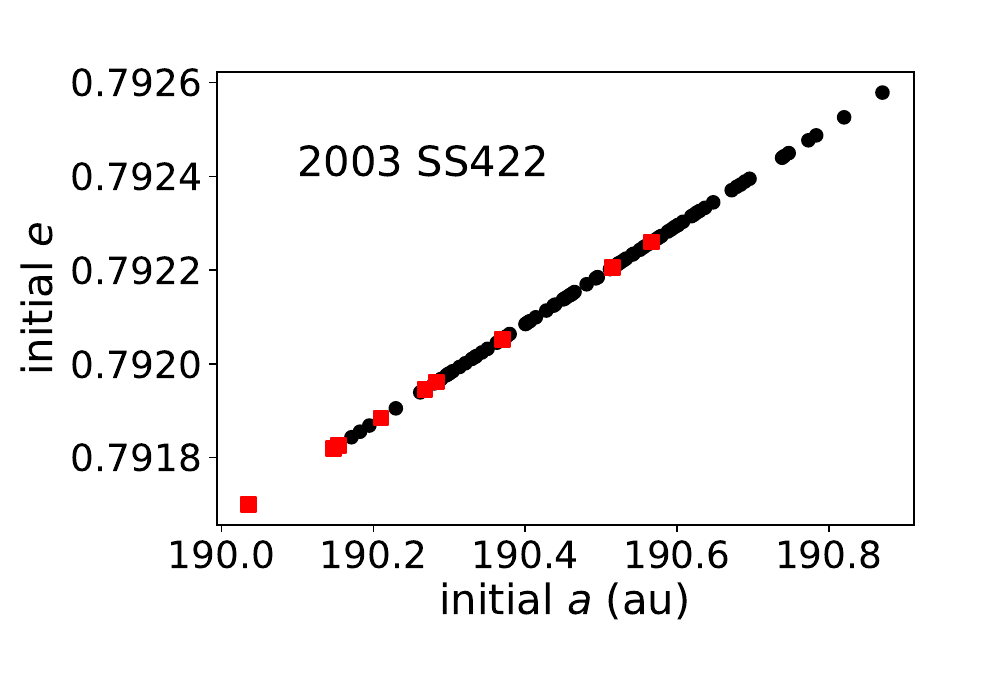}
    \caption{Distribution of initial $a$ and $e$ for 100 orbits sampled from the JPL orbit-fit and covariance matrix for TNO 2003 SS422. Black points show orbits classified as resonant by our machine learning classifier, and red squares show orbits classified as scattering. 
    The classifier predicts that 91\% of the clones are resonant, which we confirmed by visual inspection of the resonant angle $\phi=16\lambda - \lambda_N - 15\varpi$. 
    We can thus assign this TNO a 91\% probability of being in Neptune's 16:1 resonance.}
    \label{f:clones}
\end{figure}

We also anticipate more robust classifications of individual TNOs by making it possible to classify large numbers of clones sampling an individual TNO's orbit-fit uncertainties.
We show an example in Figure~\ref{f:clones}, where we use our classifier to conclude that there is a 91\% chance that TNO 2003 SS422 is currently in Neptune's 16:1 resonance based on 100 clones sampled from JPL's best-fit orbit and covariance matrix\footnote{Taken from \url{https://ssd.jpl.nasa.gov/tools/sbdb\_lookup.html#/?sstr=2003\%20SS422} for their 2022-Aug-17 21:39:48 orbit solution.}. 
The ability to assign a probability to this TNO's resonant classification is a significant improvement over noting only a `secure' or `insecure' status using the previous 3-clone approach.

Our classifier can also be used to classify synthetic TNOs, such as those produced by computer simulations of models of the early evolution of the outer solar system (see, e.g., \citealt{Kaib:2016,Pike:2017,Lawler:2019,Huang:2022,Nesvorny:2023} for recent examples of such simulations) in a way that is fully consistent with the classification of real TNOs.
This will enable much more robust comparisons between models and observations.

We note that the rate of fully `incorrect' classifications given by the classifier might be even lower than stated above, depending on how dynamically important very high-order resonances really are.
About half of the completely incorrect classifications are of particles that were predicted to be classical or detached but are actually librating in very high-order ($q>10$) resonances.
However, as discussed in Section~\ref{ss:resonances} and Figure~\ref{f:high-order-res}, it is not clear that the perturbations from these very high-order resonances meaningfully affect the long-term orbital evolution of these orbits.
Additional investigations are needed to explore which high order resonant angles reflect substantial perturbations and which might represent mere coincidental libration.
Machine learning might be a promising approach for finding the limits of Neptune's resonances as well as exploring the fuzzy boundaries between the scattering and detached populations and amongst the resonant/near-resonant/non-resonant populations.

\textcolor{black}{While we anticipate that our existing classifier is robust enough for TNOs in the LSST era, there are still improvements that could be made.
As noted in Section~\ref{ss:performance}, some data features are more influential than others in determining how well the classifier performs compared to a human as well as how often the classifier provides the undesirable dynamically irrelevant predictions. 
We were able to determine that excluding any of our 227 data features did not improve our classifier, however we have not robustly measured the trade off between accuracy and all possible combinations of those features. 
The computational cost of calculating all 227 features is relatively small but not trivial for the very large sets of particles we anticipate classifying in the LSST era.
A more extensive analysis to determine a potentially smaller subset of those features that yields acceptably accurate classifications could improve the efficiency of the classifier.}

Machine learning approaches will be critical to classifying the expected order of magnitude larger number of TNO discoveries over the next decade, and they are well-suited to tasks such as dynamical classification.
However, we note that supervised classifiers such as ours are only as good as their training set and data features.
If we discover TNOs whose dynamics are not already represented in the training set and/or whose orbital evolution is not well-characterized by the chosen data features, they will not be reliably classified. 
One might expect that any such novel dynamics would most likely be found for TNOs discovered in presently-unobserved regions of $a$-$e$-$i$ space. 
Such unusual discoveries will hopefully thus be easily spotted within the large dataset and flagged for human-driven dynamical analyses.
As we discover new TNOs, care will need to be taken to continue updating machine learning training sets and/or approaches to ensure accurate classifications. 
Future unsupervised machine learning tools could also be fruitful for guiding improved classification schemes as the population of known TNOs increases and we are able to more cleanly divide them into additional categories.

\section*{Acknowledgements}
We thank Evgeny Smirnov for a careful review that improved the chapter. 
This work was supported by NASA grant 80NSSC23K0886. KV acknowledges additional support from NASA (grants 80NSSC22K0512, 80NSSC21K0376, and 80NSSC19K0785) and the Preparing for Astrophysics with LSST Program funded by the Heising Simons Foundation (grant 2021-2975).

\section*{Code Availability}
The training set and code associated with this chapter are available on GitHub \url{https://github.com/solar-system-ml/book/tree/main/docs/chapter7}. 
The tools to initialize TNO integrations for the the machine learning classifier are available as part of the Small Body Dynamics Tool (SBDynT) \url{https://github.com/small-body-dynamics/SBDynT}. Future iterations of the TNO classifier will be available within SBDynT.

\section{Appendix--synthetic TNO orbit generation}\label{a:synthetic}

To enhance the scattering population training set, we randomly sampled semimajor axes from 52-200~au, perihelion distances from 25-43~au, inclinations from 0-45$^\circ$, and randomized the other orbital angles; we also generated a set of scattering TNOs with isotropic inclinations to double the small sample of retrograde orbits in the training set. 
These orbits were integrated in the same manner as the real TNO orbits and examined by eye to label them as scattering, detached, or resonant. 
From this set, we selected 383 scattering particles with a range of scattering behavior from weak scattering to very rapid scattering.
This dataset also included resonant and detached particles, contributing 126 additional high-$a$ resonant particles and 76 detached particles.

To increase the variety of resonant particles in the training set, we started with a list of every $p$:$q$ MMRs with $q\ge5$ already identified from the observed TNO clones and generated new orbits within a few hundredths of an au of each resonance center over a range of eccentricities similar to the overall observed population (ranging from $e\approx0-0.35$ for resonances in the classical belt region and $q\approx30-40$ in the more distant populations).
For eccentricity-type resonances, we sampled inclinations up to $\sim45^{\circ}$, chose a random longitude of ascending node and mean anomaly, then set the longitude of perihelion such that the initial value of the resonant angle would be in the range $140-220^{\circ}$, near the expected libration center, though this did not guarantee libration. 
When examining the evolution of these particles, we plotted both eccentricity and mixed-type resonance angles and labeled the resulting resonant particles appropriately. 
This set of integrations produced 4529 resonant particles, of which 626 were in mixed argument resonances. 
From the non-resonant particles, we added 275 scattering particles, 299 detached particles, and 28 classical particles (all in the outer classical belt population; we did not include any additional main belt classical particles as those require more time-intensive investigation to be sure they are non-resonant and the observed population already provides a wealth of training set data). 

Finally, to increase the number of mixed-argument resonant particles, we ran a set of orbits near every resonance with observed mixed-argument libration targeting inclination ranges seen to librate and choosing initial orbital angles to place the mixed-e-i resonant argument near the expected center of libration.  
These simulations added 451 particles exhibiting mixed argument libration to the training set.

These training set additions are not meant to be exhaustive, but rather to fill out obvious weak spots in the observed dataset without an overwhelming amount of manual classification.
The most difficult task still left un-done would be further expanding the detached population training set. 
When populating the high-$a$, high-$q$ orbital space expected for detached TNOs, it is actually very difficult to avoid resonances as Neptune's MMRs become stronger at high-$q$/low-$e$ (see \citealt{Volk:2022}). 
The resonant librations of the high-order distant resonances can be very subtle and difficult to detect even by eye, so the only way to be sure to have a non-resonant detached object is to plot many, many possible resonance angles to check. 
This is very time consuming, which is why our additions to the detached training set were more limited than for the targeted resonant populations and scattering populations.